\documentclass[submission]{eptcs}
\providecommand{\event}{ICLP DC 2021} 
\usepackage[utf8]{inputenc}

\title{Quantitative and Stream Extensions of Answer Set Programming}
\author{Rafael Kiesel
\institute{TU Vienna, Austria}
\email{rafael.kiesel@web.de}
}
\def\titlerunning{ASP Extensions}
\def\authorrunning{R. Kiesel}

\usepackage{tikz}
\usetikzlibrary{arrows.meta}
\usetikzlibrary{positioning,shapes,shadows,arrows}
\tikzstyle{abstract}=[rectangle, draw=black, rounded corners, text centered, anchor=north, text=black]
\usepackage{xcolor}
\colorlet{ampelOrange}{yellow!70!orange}
\usepackage{array}
\newcolumntype{P}[1]{>{\centering\arraybackslash}p{#1}}

\usepackage{enumitem}

\usepackage{amsmath}
\usepackage{amsthm}
\usepackage{amssymb}
\usepackage{mathtools}
\usepackage{times}
\usepackage{bbm}
\usepackage{nicefrac}
\usepackage[noabbrev]{cleveref} 
\crefname{equation}{}{}
\usepackage{bm}

\usepackage{todonotes}
\usepackage{xspace}
\usepackage{color,soul}
\usepackage{verbatim}

\usepackage{wrapfig}

\newcommand{\rem}[1]{\ensuremath{{\rm \texttt{#1}}}}
\newcommand{\LPMLN}{$\text{LP}^{\text{MLN}}$}

\newcommand{\splus}{\ensuremath{\bm{+}}}
\newcommand{\stimes}{\ensuremath{*}}

\usepackage{pdfpages}

\newcommand{\aspmc}{{\small\textsf{aspmc}}\xspace}

\begin{document}
\maketitle
\begin{abstract}
Answer Set Programming has separately been extended with constraints, to the streaming domain, and with capabilities to reason over the quantities associated with answer sets. We propose the introduction and analysis of a general framework that incorporates all three directions of extension by exploiting the strengths of Here-and-There Logic and Weighted Logic.
\end{abstract}

\section{Introduction}

While propositional Answer Set Programming (ASP) is already NP-hard and therefore powerful enough to express many challenging problems, their specification can be tedious and complicated. Further, there are relevant problems that require higher expressivity or reasoning over data that changes with time. This and the practical usage of ASP gave rise to a need for a simpler, more expressive, and more concise specification language~\cite{alviano2018aggregates,dell2003aggregate}. Thus, ASP was extended in multiple directions. We focus on the following ones:
\begin{enumerate}
\topsep=0pt
\itemsep=0pt
    \item Time Domain (TD): In \cite{beck2018lars} ASP-semantics were combined with a temporal context resulting in the Logic-based framework for Analytic Reasoning over Streams (LARS). Here, interpretations assign possibly different sets of facts to time points. Accordingly, the input language was extended with operators like $\Diamond$, corresponding to existential quantification over time points. Another temporal extension of ASP is Temporal Equilibrium Logic (TEL) \cite{cabalar2018temporal}.
    \item Quantitative Reasoning over Models (QM): Given a program we may not only be interested in its answer sets but also in reasoning with quantities associated with them. Commonly this includes:
        \begin{enumerate}
        \topsep=0pt
        \itemsep=0pt
            \item[2.1] Probabilities of Models \cite{baral2009probabilistic,lee2017lpmln,nickles2015system,de2007problog}
            \item[2.2] Preferences over Models \cite{lee2017lpmln,buccafurri1997strong}
            \item[2.3] Weighted Model Counting \cite{kimmig2011algebraic}
        \end{enumerate}
    \item Quantitative Constraints (QC): In order to simplify the specification language of ASP and improving its succinctness different constraints were introduced as basic language elements, among them:
        \begin{enumerate}
        \topsep=0pt
        \itemsep=0pt
            \item[3.1] Aggregates \cite{ferraris2011logic,dell2003aggregate}
            \item[3.2] Weight Constraints \cite{niemela1999stable}
            \item[3.3] Arithmetic Operations \cite{lierler2014relating}
            \item[3.4] Guessing in Rule-Heads \cite{niemela1999stable,lierler2014relating}
        \end{enumerate}
        \label{enum:quant_cats}
\end{enumerate}
This significantly enhanced the expressive power of programs and facilitated their specification. Naturally, we would like to use combinations of these directions. A current area that would benefit highly is that of traffic routing: Decisions need to be made based on dynamic temporal information ($\rightarrow$ TD), quantitative information about the traffic ($\rightarrow$ QC) and these decisions should be optimized in such a way that the waiting times are small ($\rightarrow$ QM).

\section{Problem Statement}
\label{sec:problem}
We are interested in a general framework which combines TD, QM, and QC, that is, we want to
\begin{center}
\textbf{Find and analyze a general framework that allows for succinct specifications and reasoning over answer sets in a streaming context.}
\end{center}
When \emph{finding} such a general framework we are faced with the following two main challenges. 

First, we have to decide how to lift the quantitative features to the temporal domain \emph{in a structured way}. There are many extensions introducing QM or QC and while quantitative extensions within a category share a purpose they tackle different \emph{sub}-problems. Furthermore, there are sometimes even multiple approaches for just one sub-problem that are defined independently. Already their relation and especially their combination is not always discussed, or only in post-hoc analysis~\cite{lee2017lpmln}. It follows that we need to find a reasonable way to capture the \emph{combined} capabilities introduced by the different extensions in a preferably \emph{uniform} manner. Reasonable here meaning a strategy different from simply taking an unstructured ``union'' of the different extensions. On the other hand, we need to avoid restricting ourselves to a significantly less powerful fragment.

Second, we need to incorporate the temporal domain. For this purpose, one can not simply reuse previous definitions in a straight forward way. To illustrate this, consider an aggregate expression of the form $\rem{sum}\{X : p(X)\}$ over a temporal domain, where $p(x)$ may hold at some time point but not at a different one. Here, we need to be able to specify whether the sum should consider only the values $x$ such that $p(x)$ holds at the current time point or those for which there exists a time point such that $p(x)$ holds. Furthermore, in the latter case, we need to differentiate whether the same value should be counted only once or according to its multiplicity. Similar problems occur also with other capabilities.


Given that we find an appropriate framework, we further need to \emph{analyze} it, in order to make it practically useful. An implementation and its application to real world use cases require an investigation of efficiently solvable and safe fragments. 

Detailed research questions concerning the creation and analysis of such a framework are given after a discussion of our methodology.

\section{Approach}
\label{sec:app}
The QC and QM extensions are mostly of a \emph{quantitative} nature, i.e.\ they involve some form of calculation. In order to capture possible computations abstractly, in an algebraic manner, we consider semirings.

A semiring $\mathcal{R}$ is an algebraic structure of the form $\mathcal{R} = (R, \oplus, \otimes, e_{\oplus}, e_{\otimes})$ where $\oplus$ and $\otimes$ are addition and multiplication with neutral elements $e_{\oplus}$ and $e_{\otimes}$ respectively. Using semirings one can capture various modes of calculation ranging from the the tropical semiring $\mathcal{R}_{\rem{trop}} = ([0, \infty], \min, +, \infty, 0)$ to the well known semiring over the real numbers $\mathbb{R} = (\mathbb{R}, +, \cdot, 0, 1)$. Semiring have previously been successfully used to capture different semantics in a uniform syntax, by parameterising definitions with a semiring. Bistarelli et al.\ \cite{bistarelli1997semiring} used them to define constraint semantics and Kimming et al.\ \cite{kimmig2011algebraic} provided a rule language with parameterised calculation. 

Notably, \emph{Weighted Logic} introduced by Droste and Gastin \cite{droste2007weighted} connects calculations in semirings and logical formulas. So called \emph{weighted formulas} like 
\[
\alpha = 15 \wedge \rem{Circus} \vee 20 \wedge \rem{Restaurant}
\]
can contain weights from a semiring $\mathcal{R}$ in addition to logical formulas. This allows the specification of calculations, which depend on the truth of logical formulas, by interpreting disjunctive connectives as addition $\oplus$ and conjunctive connectives as multiplication $\otimes$. In line with this, the neutral elements $e_{\oplus}$ and $e_{\otimes}$ replace falsum ($\bot$) and truth ($\top$). We slightly deviate from Droste and Gastin's original definition and use the more intuitive notation for the same formula
\[
\alpha = 15 \stimes \rem{Circus} \splus 20 \stimes \rem{Restaurant}
\]

Over the semiring $\mathbb{R}$ the weighted formula $\alpha$ encodes the amount of money spent on an evening depending on which places are visited. When $\rem{Circus}$ is false and $\rem{Restaurant}$ is true, its semantics is
$
15 \cdot 0 + 20 \cdot 1 = 20.
$
Over the tropical semiring $\mathcal{R}_{\rem{trop}}$ it encodes the price of the cheapest possible activity. In the same situation its semantics is
$
\min(15 + \infty, 20 + 0) = 20.
$

For QM it is sufficient to use the classical weighted semantics. When it comes to QC it is necessary to define answer set semantics for weighted formulas, since they are part of the program and influence which atomic formulas are asserted in answer sets. 


Since  are often hard to 
For this, we employ Here-and-There (HT) Logic \cite{pearce2004towards}. It can be seen as a fragment of intuitionistic logic with two worlds Here and There, with one interpretation $\mathcal{I}_{H}$ and $\mathcal{I}_T$ each such that when an atom holds Here then it also holds There. The semantics satisfies that given a program $\Pi$ an HT-interpretation $\langle \mathcal{I}_H, \mathcal{I}_T\rangle$ satisfies $\Pi$ if $\mathcal{I}_T$ classically satisfies the reduct of $\Pi$ with respect to $\mathcal{I}_{H}$. Then the answer sets $\mathcal{I}$ of $\Pi$ simply correspond to those interpretations $\mathcal{I}$ such that $\langle \mathcal{I}, \mathcal{I}\rangle$ satisfies $\Pi$ but for any subset $\mathcal{I}' \subsetneq \mathcal{I}$ it holds that $\langle \mathcal{I}', \mathcal{I}\rangle$ does not satisfy $\Pi$. 

It was shown that this definition of answer sets is equivalent to the typical reduct-based definitions~\cite{lifschitz2008answer,gelfond2014vicious} and can be combined well with other logics like first-order logic \cite{10.1007/978-3-540-89982-2_46} or temporal logic \cite{cabalar2018temporal} giving them an answer set semantics.


\section{Research Plan \& Progress}
Weighted Logic combines Boolean logic and algebraic expressions leading to the possibility to specify calculations dependent on the satisfaction of logical formulas with ease. In Weighted HT Logic, a combination of Weighted Logic and HT Logic, this dependence follows the intuition of answer set semantics. On the one hand, Weighted Logic allows many different modes of calculation due to the variability of the underlying semiring. On the other hand, its syntax and semantics is homogeneous for each of those modes. This makes Weighted (HT) Logic a promising approach to formalize many quantitative extensions of ASP for QM and QC using a common homogeneous language.

Furthermore, Weighted Logic is rather generic: for many two-valued logics it is possible to define a weighted version, by using the intuition that conjunction is multiplication and disjunction is addition. This facilitates the combination with the temporal domain.

We aim to find a general framework using Weighted (HT) Logic. Thus, we consider the following research questions:

\paragraph{Generality \& Faithfulness}
\begin{enumerate}[leftmargin=*, widest=(Q1)]
    \item[(Q1)] Can we use Weighted Logic for General Quantitative Reasoning over Models?
    \item[(Q2)] Can we use Weighted HT Logic for General Quantitative Constraints?
    \item[(Q3)] Can we capture temporal aspects using Weighted (HT) Logic?
\end{enumerate}
\noindent The first concern here is \emph{Generality}, i.e. whether the capabilities of previous extensions are subsumed to a reasonable degree. The second is \emph{Faithfulness}, i.e. if the capabilities in our framework correctly capture the intended semantics. Q1 and Q2 focus on QM and QC respectively, whereas Q3 asks whether lifting the combination of ASP and Weighted (HT) Logic to LARS/TEL sufficiently captures the temporal aspects of TD. 

Q1 has been answered positively in \cite{DBLP:conf/ecai/EiterK20}. We considered Weighted LARS (wLARS) formulas as a means of extracting quantities from answer streams of LARS programs. This lead to a general framework for the QM extensions. wLARS formulas can be used for probabilistic reasoning, preferential reasoning and weighted model counting, respectively via normalization, optimization and aggregation over all answer streams of a program. We showed how three prominent extensions of logic programming with QM capabilities of different natures, namely P-log \cite{baral2009probabilistic}, (a)Problog \cite{de2007problog,kimmig2011algebraic} and $\text{LP}^{\text{MLN}}$ \cite{lee2017lpmln} can be faithfully expressed using wLARS. Apart from that, we considered preferential reasoning with wLARS in more detail and demonstrated that the temporal aspects that need to be considered when obtaining quantities are also covered by wLARS. This is a first step towards answering Q3 positively.

Secondly, Q2 has been answered positively in \cite{DBLP:journals/tplp/EiterK20}. In order to achieve a similar level of generality for QC, we defined First-Order Weighted HT Logic. Here, the ``first-order quantifiers'' naturally correspond to aggregation. Weighted formulas were used to define \emph{algebraic constraints} $k \sim_{\mathcal{R}} \alpha$, which are (in)equations between a value $k$ and a weighted formula $\alpha$ over the semiring $\mathcal{R}$. They can be used to assert (in rule-heads) or check (in rule-bodies) whether an interpretation satisfies a restriction on some quantity (specified by $\alpha$). Taking our previous example formula $\alpha$, the algebraic constraint 
$
30 >_{\mathbb{R}} 15 \stimes \rem{Circus} \splus 20 \stimes \rem{Restaurant}
$ 
is satisfied whenever a budget of 30 is sufficient to pay for the places visited. 

We found that all the identified capabilities - aggregation, weight constraints,  arithmetic operations, ... - are faithfully incorporated in our framework, with only mild practical restrictions. We even show that there are some new capabilities, like so called "minimized choice constraints", that have not been considered before.

Therefore, the question of generality and faithfulness has already been answered positively. Only the lifting along TD using LARS may cause a problem. While there is an HT Logic for LARS, it leads to a different semantics. It remains to be seen whether this can be fixed. Otherwise, TEL, which is defined using HT Logic and adds operators for reasoning over streams, is a promising alternative.
\paragraph{Theoretical Analysis}
\begin{enumerate}[leftmargin=*, widest=(Q1)]
    \item[(Q4)] Which properties of programs carry over and which new ones can be found?
    \item[(Q5)] What does the complexity of reasoning depend on?
\end{enumerate}
The following properties are known to be of interest:
\begin{itemize}
\topsep=0pt
\itemsep=0pt
    \item \textbf{Safety}, a property necessary for ASP with variables. Considered, inter alia, in \cite{DBLP:journals/tplp/GebserHKLS15,DBLP:conf/iclp/Lifschitz16,DBLP:conf/lpnmr/CabalarPV09}
    \item \textbf{Finite Groundability}, a stronger restriction than safety that entails that programs can be evaluated over a finite domain. Considered, inter alia, in \cite{lierler2009one,efkr2013-aaai,bartholomew2010decidable}
    \item \textbf{Modularity}, a property facilitating the parallel evaluation of answer set programs. Considered, inter alia, in \cite{lifschitz1994splitting,ferraris2011logic}
    \item \textbf{Program Equivalence}, a property used in the simplification of answer set programs. Considered, inter alia, in \cite{DBLP:journals/ijar/NievesL15,lifschitz2001strongly,beck2016equivalent}
\end{itemize}
Safety and finite groundability are essential for practical usability. Unsafe programs may have unintuitive semantics. Finite Groundability ensures that one can reduce reasoning tasks for programs with variables to the corresponding reasoning tasks for programs without variables. 

We already introduced safety conditions for programs with algebraic constraints~\cite{DBLP:journals/tplp/EiterK20}. Furthermore, we combined and transferred the concepts of domain restrictedness from \cite{niemela1999stable} and argument restrictedness from \cite{lierler2009one} to our framework, resulting in a class of finitely ground programs, for which membership can be decided tractably. Also, regarding program equivalence, we showed that two programs with algebraic constraints are strongly equivalent iff their HT-models coincide~\cite{DBLP:journals/tplp/EiterK20}.

Modularity may still be considered in-depth later on.

As for complexity, our preliminary results show strong influence by two parameters. The first is which semirings are allowed: Depending on the semiring a single addition can be incomputable or constant. Already in~\cite{DBLP:conf/ecai/EiterK20,DBLP:journals/tplp/EiterK20} we gave a preliminary consideration of the complexity of the respective introduced reasoning tasks. There, we identified some sufficient conditions on semirings for efficient calculations, which allowed us to provide upperbounds on the complexity. 

However, we noticed that even under the chosen conditions the choice of semiring leads to a lot of variation in the complexity: In the context of QM the problem of Algebraic Answer Set Counting (AASC) (i.e.\ counting answer sets, which have a weight over a semiring) can be \textsc{NP}, \textsc{\#P} or \textsc{OptP}-complete depending on the choice of the semiring. This motivated a more in depth consideration of the class of Sum-Of-Products problems over semirings in~\cite{DBLP:conf/aaai/EiterK21}. Apart from AASC, also many other problems (viz.\ \cite{kimmig2011algebraic,bistarelli1997semiring,DBLP:journals/tplp/EiterK20}) fall into this class of problems. We introduced \textsc{NP($\mathcal{R}$)} a generalization of \textsc{NP} parameterized with a semiring $\mathcal{R}$ together with a prototypical complete problem \textsc{SAT($\mathcal{R}$)} and showed that AASC and many other problems are \textsc{NP($\mathcal{R}$)}-complete. Furthermore, we analyzed the relation of \textsc{NP($\mathcal{R}$)} to classical complexity classes in dependence on the semiring $\mathcal{R}$ and the properties satisfied by it.

The second parameter is the allowed language fragment. We can tune both the expressiveness and the restrictions on variable occurrences. We plan to consider different conditions for finite groundability and consider the complexity entailed by using them. 

\paragraph{Empirical Analysis}
\begin{enumerate}[leftmargin=*, widest=(Q1)]
    \item[(Q6)] To what extent can we efficiently implement our framework?
    \item[(Q7)] How can our framework be employed in real world applications?
\end{enumerate}
We obtained promising preliminary results regarding Q6 by considering AASC, i.e., Weighted LARS restricted to non-disjunctive ASP~\cite{eiter2021treewidth}. Here, we introduced a new way to translate a program $\Pi$ into propositional formula $\phi$ such that the treewidth of $\phi$ is bounded by $kc$, where $k$ is the treewidth of $\Pi$ and $c$ is \emph{component-boosted backdoor size}, a novel parameter that measures the cyclicity of the dependency graph of $\Pi$. In addition to that, our translation satisfies that the (weighted/algebraic) answer set count of $\Pi$ is equal to the (weighted/algebraic) model count of $\phi$. Our experimental results showed that our prototype implementation \aspmc\footnote{available at \href{https://github.com/hmarkus/asp2sat}{https://github.com/hmarkus/asp2sat} (open source).} outperforms the current standard tool called Problog~\cite{de2007problog} on a standard benchmark set from \cite{tsamoura2020beyond}. We thus argue that \aspmc provides an efficient implementation for QM. It remains to be seen in the future, how this implementation can be extended to the temporal domain, i.e., Weighted LARS.

For QC an implementation is further in the future and it seems likely that we will focus on extending \aspmc to also handle the temporal domain. In the same step, we also plan to apply it in a real world application in the context of real time traffic regulation in order to answer Q7.


\section{State of the Art}
To the best of my knowledge, the idea of using Weighted Logic as a means of adding general quantitative reasoning capabilities to ASP is novel. There has however already been research useful for expressing multiple quantitative extensions of ASP in a common terminology. In the following, we go over different quantitative extensions of ASP aiming at generality.
\paragraph{Hybrid ASP \cite{cabalar2020uniform}}
The authors defined an extension of HT Logic that includes general constraints and multi-valued interpretations 
providing a non-monotonic integration of constraint processing into ASP. Since the approach allows arbitrary constraints it captures QC extensions; however, as a consequence, the syntax of constraints (except over the reals) is left open, and specific constraint expressions like aggregates rely on extra definitions.

\paragraph{Nested Expressions~\cite{ferraris2011logic}}
The formalism allows the nesting of constraints on aggregates. The only restriction on aggregates is that they have to be over the reals. Three QC extensions are partly subsumed by the approach. 
\paragraph{\LPMLN~\cite{lee2017lpmln}}
Lee and Yang's extension of logic programming with probabilistic reasoning has been shown to subsume two other QM extensions, namely ASP with weak constraints and P-log. This shows that \LPMLN captures multiple capabilities along QM. It however neglects problems that are not over the reals.

\paragraph{Problog~\cite{de2007problog}}
Problog is an implementation for probabilistic logic programming. It even also allows \emph{algebraic} answer set counting, however, it only allows a restricted fragment of ASP in the input language.

\paragraph{telingo~\cite{cabalar2018temporal}}
Telingo is a solver from the potassco family, built for a fragment of TEL, a logic for temporal reasoning under ASP semantics. It already combines TD and QC, however the temporal aspects are not fully incorporated in the quantitative constraints. It may be useful as a basis of an implementation for our work.

\section*{Acknowledgments} The author would like to thank Thomas Eiter for supervising his PhD research activities. This work has been supported by FWF project W1255-N23. 
\bibliographystyle{eptcs}
\bibliography{bib.bib}




\end{document}